\documentclass[12pt]{article}
\usepackage[dvips]{graphicx}
\usepackage{amssymb}
\begin{document}
%\twocolumn{}
\begin{center}
{ \LARGE {\bf A New form of Kr\"{o}necker product-Lax pair and coupled systems
\\\
 } }
\end{center}
\vskip 0pt
\begin{center}
{\it {\large $Arindam\hskip 2 pt Chakraborty $\\{Department
of Physics,Heritage Institute of Technology,Kolkata,India}  $A. Roy \hskip 3 ptChowdhury
$\footnote {(Corresponding author)e-mail: asesh\_r@yahoo.com}  }\\
\it{High Energy Physics Division, Department of Physics ,
                             Jadavpur University,\\
                              Kolkata - 700032,
                                    India}  }
\end{center}

\vskip 20pt
\begin{center}
{\bf Abstract}
\end{center}

\par A new method is proposed to generate nonlinear integrable systems by starting with existing Lax pair and a new form of Kr\"{o}necker product. It is observed that such equation can be generated with the help of a Hamiltonian structure. Actually they are all bi-Hamiltonian. A few explicit form of such equations are given also.

 \par PACS Number(s):  05.45.Pq, 05.45.Ac, 05.45.-a.

\section{Introduction}
Extending the class of integrable systems is one of the most important aspects of research in the field of solitons. Regarding this direction the pioneering work was done by Fucsshteiner${[1]}$, who showed how starting with a system and its linearized one or with its perturbed expression one can get new coupled systems. In these analysis an important role was played by the hereditary operators${[2]}$ and corresponding Lie symmetries${[3]}$. An application with some modification was done by Zang et. al.${[4]}$ Some important work was done for an infinite dimensional integrable system by Bacocai and Zhuqan${[5]}$. On the other hand coupling of commutator pairs as a new means of generating coupled integrable systems was done by Zhang and Tau${[6]}$ and their method basically depends on the decomposition of a Lie algebra on its various subalgebras. But one cannot forget the methodology of Whalquist and Estabrook${[7]}$ using the formalism of prolongation structure to generate a Lax pair. Such an approach was used by Roychowdhury et. al.${[8]}$ and Yu Fa Jun${[9]}$. Also It can be recalled that any integrable hierarchy contains both positive and negative order equation sets${[10]}$. An important equation is the sine-Gordin Equation${[11]}$. The coupling of negative order equations was done by Li Li and F. Yu${[12]}$. In this respect one cannot overlook the case of $(2+1)$-dimensional systems. Some important work has already been done by Feng et. al.${[13]}$
\par In the present communication we have proposed a new form of Kr\"{o}necker product Lax operator with the help of original space-time pair along with some auxiliary operators which involve projection type operators and their properties. This leads us to a new class of integrable coupled systems. Our method can also be extended to incorporate more than two sets of integrable systems and various forms of couplings. Each such hierarchy is bi-Hamiltonian.
\section{Formulation}
Let us consider two arbitrary integrable systems whose space parts of Lax pairs are
\begin{eqnarray}
\psi_x=U_1\psi\nonumber\\
\phi_x=U_2\phi
\end{eqnarray}
corresponding time parts are
\begin{eqnarray}
\psi_t=V_1\psi\nonumber\\
\phi_t=V_2\phi
\end{eqnarray}
Let us now construct Kr\"{o}necker product matrix;
\begin{eqnarray}
U=M_1\otimes U_1+M_2\otimes U_2\nonumber\\
V=M_1\otimes V_1+M_2\otimes V_2
\end{eqnarray}
and consider the compatibility of
\begin{eqnarray}
\chi_x=U\chi\nonumber\\
\chi_t=V\chi
\end{eqnarray}
Choosing $M_1=\bf{1}$ and $M_2=P_{-}$ with $P^2_{-}=\bf{1}$ leads to
\par Case-1:
\begin{eqnarray}
U_{1t}-V_{1x}+[U_1,V_1]+[U_2,V_2]=0\nonumber\\
U_{2t}-V_{2x}+[U_1,V_2]+[U_2,V_1]=0
\end{eqnarray}
On the other hand if we choose $M_1=\bf{1}$ and $M_2=P$ with $P^2_{-}=P$  the the compatibility leads to
\par Case-2:
\begin{eqnarray}
U_{1t}-V_{1x}+[U_1,V_1]&=&0\nonumber\\
U_{2t}-V_{2x}+[U_1,V_2]+[U_2,V_1]+[U_2,V_2]&=&0
\end{eqnarray}
\par Case-3:
\begin{eqnarray}
U=\textbf{1} \otimes U_1+P_0\otimes U_2\nonumber\\
V=\textbf{1}\otimes V_1+P_0\otimes V_2
\end{eqnarray}
with $P_0^2=-P_0$
In this situation we get
\begin{eqnarray}
U_{1t}-V_{1x}+[U_1,V_1]=0\nonumber\\
U_{2t}-V_{2x}+[U_1,V_2]+[U_2,V_1]-[U_2,V_2]=0
\end{eqnarray}
\par Case-4:
\begin{eqnarray}
U=\textbf{1} \otimes U_1+P_1\otimes U_2\nonumber\\
V=\textbf{1}\otimes V_1+P_1\otimes V_2
\end{eqnarray}
with $P_1^2=-\bf{1}$ then the resulting system reads
\begin{eqnarray}
U_{1t}-V_{1x}+[U_1,V_1]-[U_2,V_2]=0\nonumber\\
U_{2t}-V_{2x}+[U_1,V_2]-[U_2,V_1]=0
\end{eqnarray}
\par $\bf{Example-I:}$
\par Considering a typical AKNS system the loop algebra $su(2)\otimes\lambda^n$  and taking
\begin{eqnarray}
U_1&=&-2\alpha(1)+q\beta_1(0)+r\beta_2(0)\nonumber\\
U_2&=&\omega_1\beta_1(0)+\omega_2\beta_2(0)
\end{eqnarray}
where the basic commutation rules are
\begin{eqnarray}
[\alpha(m),\beta_1(n)]=\beta_1(m+n)\nonumber\\
\end{eqnarray}
\begin{eqnarray}
[\alpha(m),\beta_2(n)]=-\beta_2(m+n)\nonumber\\
\end{eqnarray}
\begin{eqnarray}
[\beta_1(m),\beta_2(n)]=2\alpha(m+n)
\end{eqnarray}
with $\alpha(j)=\frac{1}{2}\sigma_3\otimes\lambda^j$, $\beta_1(j)=\sigma_{+}\otimes\lambda^j$, $\beta_2(j)=\sigma_{-}\otimes\lambda^j$.
In view of case-1 if we choose
\begin{eqnarray}
V_1&=&\sum_{j}[A_j\alpha(-j)+B_{j}\beta_1(-j)+C_{j}\beta_2(-j)]\nonumber\\
V_2&=&\sum_{j}[Q_j\alpha(-j)+R_{j}\beta_1(-j)+S_{j}\beta_2(-j)]
\end{eqnarray}
the compatibility condition yields the following relations
\begin{eqnarray}
A_{mx}&=&2qC_m-2rB_m+2\omega_1S_m-2\omega_2R_m\nonumber\\
B_{mx}&=&-2B_{m+1}-q A_m-\omega_1Q_m\nonumber\\
C_{mx}&=&2C_{m+1}+rA_m+\omega_2Q_m
\end{eqnarray}
along with those given below
\begin{eqnarray}
Q_{mx}&=&2qS_m-2rR_m+2\omega_1C_m-2\omega_2B_m\nonumber\\
R_{mx}&=&-2R_{m+1}-q Q_m-\omega_1A_m\nonumber\\
S_{mx}&=&2S_{m+1}+rQ_m+\omega_2A_m
\end{eqnarray}
and the following evolution equations
\begin{eqnarray}
q_t=2B_{n+1}\nonumber\\
r_t=-2C_{n+1}\nonumber\\
\omega_{1t}=2R_{n+1}\nonumber\\
\omega_{2t}=-2S_{n+1}
\end{eqnarray}
After some tricky manipulation we can obtain the recursion relation
\begin{eqnarray}
\Lambda_{n+1}=M\Lambda_n
\end{eqnarray}
where $\Lambda_i=(B_i,C_i, R_i, S_i)^T$ and the elements of $M$ are
\begin{eqnarray}
M_{11}&=&-\frac{1}{2}\partial+q\partial^{-1}r+\omega_{1}\partial^{-1}\omega_{2}\nonumber\\
M_{12}&=&-q\partial^{-1}q-\omega_{1}\partial^{-1}\omega_{1}\nonumber\\
M_{21}&=&r\partial^{-1}r+\omega_{2}\partial^{-1}\omega_{2}\nonumber\\
M_{22}&=&\frac{1}{2}\partial-r\partial^{-1}q-\omega_{2}\partial^{-1}\omega_{1}\nonumber\\
M_{31}&=&q\partial^{-1}\omega_{2}+\omega_{1}\partial^{-1}r\nonumber\\
M_{32}&=&-q\partial^{-1}\omega_{1}-\omega_{1}\partial^{-1}q\nonumber\\
M_{41}&=&r\partial^{-1}\omega_{2}+\omega_{2}\partial^{-1}r\nonumber\\
M_{42}&=&-r\partial^{-1}\omega_{1}-\omega_{2}\partial^{-1}q\nonumber\\
M_{13}&=&q\partial^{-1}\omega_{2}+\omega_{1}\partial^{-1}r\nonumber\\
M_{14}&=&-q\partial^{-1}\omega_{1}-\omega_{1}\partial^{-1}q\nonumber\\
M_{23}&=&r\partial^{-1}\omega_{2}+\omega_{2}\partial^{-1}r\nonumber\\
M_{24}&=&-r\partial^{-1}\omega_{1}-\omega_{2}\partial^{-1}q\nonumber\\
M_{33}&=&-\frac{1}{2}\partial+q\partial^{-1}r+\omega_{1}\partial^{-1}\omega_2\nonumber\\
M_{34}&=&-q\partial^{-1}q-\omega_{1}\partial^{-1}\omega_1\nonumber\\
M_{43}&=&r\partial^{-1}r+\omega_{2}\partial^{-1}\omega_2\nonumber\\
M_{44}&=&\frac{1}{2}\partial-r\partial^{-1}q-\omega_{2}\partial^{-1}\omega_1\nonumber\\
\end{eqnarray}
With the help of $M$ one can generate a hierarchy of coupled integrable systems. As a simple example if you choose $B_0=q$, $C_0=r$, $S_0=\omega_1$ and $R_0=\omega_1$
\begin{eqnarray}
q_t&=&\frac{1}{2}q_{xx}-q(qr+\omega_1\omega_2)-\omega_1(q\omega_2+r\omega_1)\nonumber\\
r_t&=&-\frac{1}{2}r_{xx}+r(qr+\omega_1\omega_2)+\omega_2(q\omega_2+r\omega_1)\nonumber\\
\omega_{1t}&=&\frac{1}{2}\omega_{1xx}-q(q\omega_2+r\omega_1)-\omega_1(qr+\omega_1\omega_2)\nonumber\\
\omega_{2t}&=&-\frac{1}{2}\omega_{2xx}+r(q\omega_2+r\omega_1)+\omega_2(qr+\omega_1\omega_2)
\end{eqnarray}
\par $\bf{Example-II:}$
Considering case-2 for a typical Koup-Newell system with the following choice of

\begin{eqnarray}
U_1&=&-2i\alpha(2)+q\beta_1(1)+r\beta_2(1)\nonumber\\
U_2&=&u_1\beta_1(1)+u_2\beta_2(1)
\end{eqnarray}
and for the time part
\begin{eqnarray}
V_1&=&\sum_{j}[a_j\alpha(-2j)+b_{j}\beta_1(-2j-1)+c_{j}\beta_2(-2j-1)]\nonumber\\
V_2&=&\sum_{j}[d_j\alpha(-2j)+e_{j}\beta_1(-2j-1)+f_{j}\beta_2(-2j-1)]
\end{eqnarray}
Proceeding as before we arrived at
\begin{eqnarray}
\Theta_t=\partial \Pi_m
\end{eqnarray}
along with
\begin{eqnarray}
\Pi_{n+1}=M \Pi_n
\end{eqnarray}
where $\Theta_t=(q_t, r_t, u_{1t}, u_{2t})^T$ and $\Pi_i=(b_i, c_i, e_i, f_i)^T$. The matrix $M$ is given by
\begin{eqnarray}
M_{11}&=&\frac{1}{2}(i\partial+q\partial^{-1}r\partial)\nonumber\\
M_{12}&=&\frac{1}{2}q\partial^{-1}q\partial\nonumber\\
M_{21}&=&\frac{1}{2}r\partial^{-1}r\partial\nonumber\\
M_{22}&=&-\frac{1}{2}(i\partial-r\partial^{-1}q\partial)\nonumber\\
M_{31}&=&\frac{1}{2}(q+u_1)\partial^{-1}u_2\partial+\frac{1}{2}u_1\partial^{-1}r\partial\nonumber\\
M_{32}&=&\frac{1}{2}(q+u_1)\partial^{-1}u_1\partial+\frac{1}{2}u_1\partial^{-1}q\partial\nonumber\\
M_{33}&=&\frac{i}{2}\partial+\frac{1}{2}(q+u_1)\partial^{-1}(r+u_2)\partial\nonumber\\
M_{34}&=&\frac{1}{2}(q+u_1)\partial^{-1}(q+u_1)\partial\nonumber\\
M_{41}&=&\frac{1}{2}(r+u_2)\partial^{-1}u_2\partial+\frac{1}{2}u_2\partial^{-1}r\partial\nonumber\\
M_{42}&=&\frac{1}{2}(r+u_2)\partial^{-1}u_1\partial+\frac{1}{2}u_2\partial^{-1}q\partial\nonumber\\
M_{43}&=&\frac{1}{2}(r+u_2)\partial^{-1}(r+u_2)\partial\nonumber\\
M_{44}&=&-\frac{i}{2}\partial+\frac{1}{2}(r+u_2)\partial^{-1}(q+u_1)\partial\nonumber\\
\end{eqnarray}
with $M_{13}=M_{14}=M_{23}=M_{24}=0$
In this case we can also write down the first set of equations with the choice $b_0=r, c_0=q, e_0=u_2, f_0=u_1$
\begin{eqnarray}
q_t&=&\frac{i}{2}r_{xx}+\frac{1}{4}[q(r^2+q^2)]_x\nonumber\\
r_t&=&-\frac{i}{2}q_{xx}+\frac{1}{4}[r(r^2+q^2)]_x\nonumber\\
u_{1t}&=&\frac{i}{2}u_{2xx}+\frac{1}{2}[(q+u_1)(qu_1+ru_2)]_x+\frac{1}{4}[(q+u_1)(u_1^2+u_2^2)+u_1(q^2+r^2)]_x\nonumber\\
u_{2t}&=&-\frac{i}{2}u_{1xx}+\frac{1}{2}[(r+u_2)(qu_1+ru_2)]_x+\frac{1}{4}[(r+u_2)(u_1^2+u_2^2)+u_2(q^2+r^2)]_x
\end{eqnarray}
\par $\bf{Example-III:}$
We now proceed to a more elaborate situation to induce coupling between more than two nonlinear systems. Let us take;
\begin{eqnarray}
U=\textbf{1} \otimes U_1+P\otimes U_2+P^{\prime}\otimes U_3+P_{-}\otimes U_4\nonumber\\
V=\textbf{1} \otimes V_1+P\otimes V_2+P^{\prime}\otimes V_3+P_{-}\otimes V_4\nonumber\\
\end{eqnarray}
With the following properties of $\{P_i\}$'s:$P^2=P$, $P^{\prime 2}=P^{\prime}$, $PP^{\prime}=P^{\prime}P=0$, $PP_{-}=P_{-}P=P$ and $P_{-}P^{\prime}=P^{\prime}P_{-}=-P^{\prime}$
\par The relevant Lax equation
\begin{eqnarray}
U_t-V_x+[U,V]=0
\end{eqnarray}
leads to
\begin{eqnarray}
U_{1t}-V_{1x}+[U_1,V_1]+[U_4,V_4]&=&0\nonumber\\
U_{2t}-V_{2x}+[U_1,V_2]+[U_2,V_1]+[U_2,V_2]+[U_2,V_4]+[U_4,V_2]&=&0\nonumber\\
U_{3t}-V_{3x}+[U_1,V_3]+[U_3,V_1]+[U_3,V_3]-[U_3,V_4]-[U_4,V_3]&=&0\nonumber\\
U_{4t}-V_{4x}+[U_1,V_4]+[U_4,V_1]&=&0
\end{eqnarray}
where we choose
\begin{eqnarray}
U_1&=&-2\alpha(1)+q\beta_1(0)+r\beta_2(0)\nonumber\\
U_2&=&u_1\beta_1(0)+u_2\beta_2(0)\nonumber\\
U_3&=&v_1\beta_1(0)+v_2\beta_2(0)\nonumber\\
U_4&=&\omega_1\beta_1(0)+\omega_2\beta_2(0)
\end{eqnarray}
as space parts and
\begin{eqnarray}
V_1&=&\sum_j[A_j\alpha(-j)+B_j\beta_1(-j)+C_j\beta_2(-j)]\nonumber\\
V_2&=&\sum_j[D_j\alpha(-j)+E_j\beta_1(-j)+F_j\beta_2(-j)]\nonumber\\
V_3&=&\sum_j[H_j\alpha(-j)+K_j\beta_1(-j)+L_j\beta_2(-j)]\nonumber\\
V_4&=&\sum_j[Q_j\alpha(-j)+R_j\beta_1(-j)+S_j\beta_2(-j)]
\end{eqnarray}
as corresponding time parts leading to the following recursion relations
\begin{eqnarray}
A_{mx}&=&2qC_m-2rB_m+2\omega_1S_m-2\omega_2R_m\nonumber\\
B_{mx}&=&-2B_{m+1}-q A_m-\omega_1Q_m\nonumber\\
C_{mx}&=&2C_{m+1}+rA_m+\omega_2Q_m\nonumber\\
D_{mx}&=&2(q+u_1+\omega_1)F_m-2(r+u_2+\omega_2)E_m+2u_1C_m-2u_2B_m+2u_1S_m-2u_2R_m\nonumber\\
E_{mx}&=&-2E_{m+1}-(q+u_1+\omega_1) D_m-u_1(Q_m+A_m)\nonumber\\
F_{mx}&=&2F_{m+1}+(r+u_2+\omega_2)D_m+u_2(Q_m+A_m)\nonumber\\
H_{mx}&=&2(q+v_1-\omega_1)L_m-2(r+v_2-\omega_2)K_m+2v_1C_m-2v_2B_m-2v_1S_m+2v_2R_m\nonumber\\
K_{mx}&=&-2K_{m+1}-(q+v_1-\omega_1) H_m-v_1(A_m-Q_m)\nonumber\\
L_{mx}&=&2L_{m+1}+(r+v_2-\omega_2)H_m+v_2(A_m-Q_m)\nonumber\\
Q_{mx}&=&2qS_m-2rR_m+2\omega_1C_m-2\omega_2B_m\nonumber\\
R_{mx}&=&-2R_{m+1}-q Q_m-\omega_1A_m\nonumber\\
S_{mx}&=&2S_{m+1}+rQ_m+\omega_2A_m
\end{eqnarray}
An explicit form of recursion relations can be written as

\begin{eqnarray}
\Omega_{m+1}=M\Omega_m
\end{eqnarray}
where
\begin{eqnarray}
\Omega_{i}=(B_i,C_i,E_i,F_i,K_i,L_i,R_i,S_i)^T
\end{eqnarray}
The matrix $M$ has a very complicated expression which is given in the appendix. Here also a special choice of $B_0$, $C_0$ etc leads to a specific set of coupled nonlinear equations which are also given in the appendix.
\section{Hamiltonian Structure}
\par From the inception of study of integrable systems it has been observed that they can be deduced from a Hamiltonian and a symplectic operator. Actually many of them are found to be bi-Hamiltonian. An elegant method for the derivation of the Hamiltonian form was proposed by Tu${[14]}$. In this method a Lie algebra and its corresponding loop algebra $\tilde{G}=G\bigotimes C(\lambda, \lambda^{-1})$ is a Laurent polynomial in $\lambda$. Then one introduces the isospectral problem $\psi_x=U\psi$ and $\psi_t=V\psi$ where $U=R+\sum_{i=1}^p u_ie_i$ where $R$ stands for a pseudoregular element in $\tilde G$ and $\{e_i\mid i=1\dots p\}$ is a basis in $\tilde G$. $\{u_i(x,t)\mid i=1\dots p\}$ are functions called potential functions. One usually assign a degree to every element $\deg(X\bigotimes \lambda^n)=n$, $X\in G$, $\deg(R)=\alpha$ and $\deg(e_i)=\varepsilon_i$. If $\alpha\geq\epsilon$ then one can solve
\begin{eqnarray}
V_x=[U, V]
\end{eqnarray}
for $V$ as done in previous section. The next step is to choose a modified term $\Delta_n\in\tilde{G}$ so that
\begin{eqnarray}
-V_x^{(n)}+[U, V^{(n)}]=-V_{+x}^{(n)}+[U, V_{+}^{(n)}]-(\Delta_n)_x
\end{eqnarray}
where
\begin{eqnarray}
-V_{+}^{(n)}=(\lambda_nV)_{+}
\end{eqnarray}
is the positive part of $\lambda^n V$.
\par The forth step is to  use the full Lax equation to get the nonlinear system with corresponding $V_n$. Finally we seek the Hamiltonian structure by using the trace identity
\begin{eqnarray}
\frac{\delta}{\delta u}\langle{V,\frac{\partial U}{\partial \lambda}}\rangle=\left(\lambda^{-\gamma}\frac{\partial}{\partial\lambda}\lambda^{\gamma}\right)\langle{V,\frac{\partial U}{\partial u_i}}\rangle
\end{eqnarray}
with

where $\frac{\delta}{\delta u}=(\frac{\delta}{\delta u_i}\dots \frac{\delta}{\delta u_p})^T$ being the variational derivative with respect to $u=(u_1\dots u_p)$ and $\langle a,b\rangle$ stands for the matrix trace, $\langle x, y\rangle=Tr(x,y): x, y\in G$.
\par In our example-I, we get
\begin{eqnarray}
\langle{V,\frac{\partial U}{\partial \lambda}}\rangle&=&-4A\nonumber\\
\langle{V,\frac{\partial U}{\partial q}}\rangle&=&2C\nonumber\\
\langle{V,\frac{\partial U}{\partial r}}\rangle&=&2B\nonumber\\
\langle{V,\frac{\partial U}{\partial \omega_1}}\rangle&=&2S\nonumber\\
\langle{V,\frac{\partial U}{\partial \omega_2}}\rangle&=&2R
\end{eqnarray}
The trace identity yields $\gamma=-3$ and the hamiltonian structure can be written as
\begin{eqnarray}
(q,r,\omega_1, \omega_2)^T=J\left(\frac{\delta}{\delta q},\frac{\delta}{\delta r},\frac{\delta}{\delta \omega_1},\frac{\delta}{\delta \omega_2} \right)
\end{eqnarray}
\begin{eqnarray}
J= \left(\begin{array}{cccc}
   0 & 2 & 0 & 0\\
   -2 & 0 & 0 & 0\\
   0 & 0 & 0 & 2\\
   0 & 0 & -2 & 0
   \end{array} \right)
\end{eqnarray}
and
\begin{eqnarray}
H_n=\frac{2A_{n+2}}{n+4}
\end{eqnarray}
Similarly for example-II we get

\begin{eqnarray}
\langle{V,\frac{\partial U}{\partial \lambda}}\rangle&=&-4i\lambda(2a+d)+2(br+cq)+(er+bu_2)+(fq+cu_1)+(eu_2+fu_1)\nonumber\\
\langle{V,\frac{\partial U}{\partial q}}\rangle&=&\lambda(2c+f)\nonumber\\
\langle{V,\frac{\partial U}{\partial r}}\rangle&=&\lambda(2b+e)\nonumber\\
\langle{V,\frac{\partial U}{\partial u_1}}\rangle&=&\lambda(c+f)\nonumber\\
\langle{V,\frac{\partial U}{\partial u_2}}\rangle&=&\lambda(b+e)
\end{eqnarray}
so that we get
\begin{eqnarray}
(q,r,u_1, u_2)^T=J\left(\frac{\delta}{\delta q},\frac{\delta}{\delta r},\frac{\delta}{\delta u_1},\frac{\delta}{\delta u_2} \right)
\end{eqnarray}
with
$J=J_1\partial$
where
\begin{eqnarray}
J_1= \left(\begin{array}{cccc}
   0 & -2 & 0 & -1\\
   -2 & 0 & -1 & 0\\
   0 & -1 & 0 & -1\\
   -1 & 0 & -1 & 0
   \end{array} \right)
\end{eqnarray}
 and
\begin{eqnarray}
H_m=\frac{1}{2m+2}[4i(2a_{m+1}+d_{m+1})-2(rb_m+qc_m)-(rc_m+u_2b_m)-(u_1e_m+qf_m)-(u_2e_m+u_1f_m)]
\end{eqnarray}
\par Finally in example-III the situation is very complicated and we have got the following result:
\begin{eqnarray}
\langle{V,\frac{\partial U}{\partial \lambda}}\rangle&=&-2(2A+D+H)\nonumber\\
\langle{V,\frac{\partial U}{\partial q}}\rangle&=&\lambda(2C+F+L)\nonumber\\
\langle{V,\frac{\partial U}{\partial r}}\rangle&=&\lambda(2B+K+E)\nonumber\\
\langle{V,\frac{\partial U}{\partial u_1}}\rangle&=&\lambda(C+F+S)\nonumber\\
\langle{V,\frac{\partial U}{\partial u_2}}\rangle&=&\lambda(B+E+R)\nonumber\\
\langle{V,\frac{\partial U}{\partial v_1}}\rangle&=&\lambda(C+L-S)\nonumber\\
\langle{V,\frac{\partial U}{\partial v_2}}\rangle&=&\lambda(B+K-R)\nonumber\\
\langle{V,\frac{\partial U}{\partial \omega_1}}\rangle&=&\lambda(F+2S-L)\nonumber\\
\langle{V,\frac{\partial U}{\partial \omega_2}}\rangle&=&\lambda(E-K+2R)
\end{eqnarray}
leading to
\begin{eqnarray}
\left(\frac{\delta}{\delta q},\frac{\delta}{\delta r}, \frac{\delta}{\delta u_1}, \frac{\delta}{\delta u_2}, \frac{\delta}{\delta v_1}, \frac{\delta}{\delta v_2}, \frac{\delta}{\delta \omega_1}, \frac{\delta}{\delta \omega_2} \right)^T=M(B_{n+1},C_{n+1}, E_{n+1}, F_{n+1},K_{n+1}, L_{n+1}, R_{n+1}, S_{n+1} )
\end{eqnarray}
with
\begin{eqnarray}
M= \left(\begin{array}{cccccccc}
   0 & 2 & 0 & 1 & 0 & 1 & 0 & 0\\
   2 & 0 & 1 & 0 & 1 & 0 & 0 & 0\\
   0 & 1 & 0 & 1 & 0 & 0 & 0 & 1\\
   1 & 0 & 1 & 0 & 0 & 0 & 1 & 0\\
   0 & 1 & 0 & 0 & 0 & 1 & 0 & -1\\
   1 & 0 & 0 & 0 & 1 & 0 & -1 & 0\\
   0 & 0 & 0 & 1 & 0 & -1 & 0 & 2\\
   0 & 0 & 1 & 0 & -1 & 0 & 2 & 0
   \end{array} \right)
\end{eqnarray}
and the Hamiltonian
\begin{eqnarray}
H_n=\frac{1}{n+4}[4A_{n+2}+2D_{n+2}+2H_{n+2}]
\end{eqnarray}
\section{Conclusion}
In our above analysis we have shown how a new form of Kr\"{o}necker product Lax operators formed with the help of projection type operators can lead to various possible forms of new integrable systems. In each case we have shown how the method of trace identity can be used to determine its Hamiltonian structure and also the recursion operator for each heirarchy. It may be added  that the whole procedure can be extended in $(2+1)$-dimension. Such an extension will be reported in a future communication.
\section{Appendix}
\par \textbf{Elements of matrix M in equation(34)}
\begin{eqnarray}
M_{11}&=&-\frac{1}{2}\partial+q\partial^{-1}r+\omega_1\partial^{-1}\omega_2\nonumber\\
M_{12}&=&-q\partial^{-1}q-\omega_1\partial^{-1}\omega_1\nonumber\\
M_{17}&=&q\partial^{-1}\omega_2+\omega_1\partial^{-1}r\nonumber\\
M_{18}&=&-q\partial^{-1}\omega_1-\omega_1\partial^{-1}q\nonumber\\
M_{21}&=&r\partial^{-1}r+\omega_2\partial^{-1}\omega_2\nonumber\\
M_{22}&=&\frac{1}{2}\partial-r\partial^{-1}q-\omega_2\partial^{-1}\omega_1\nonumber\\
M_{27}&=&r\partial^{-1}\omega_2+\omega_2\partial^{-1}r\nonumber\\
M_{28}&=&-r\partial^{-1}\omega_1-\omega_2\partial^{-1}q\nonumber\\
M_{31}&=&(q+\omega_1+u_1)\partial^{-1}u_2+u_1\partial^{-1}(r+\omega_2)\nonumber\\
M_{32}&=&-(q+\omega_1+u_1)\partial^{-1}u_1-u_1\partial^{-1}(q+\omega_1)\nonumber\\
M_{33}&=&-\frac{1}{2}\partial+(q+\omega_1+u_1)\partial^{-1}(r+\omega_2+u_2)\nonumber\\
M_{34}&=&(q+\omega_1+u_1)\partial^{-1}(q+\omega_1+u_1)\nonumber\\
M_{37}&=&(q+\omega_1+u_1)\partial^{-1}u_2+u_1\partial^{-1}(r+\omega_2)\nonumber\\
M_{38}&=&-(q+\omega_1+u_1)\partial^{-1}u_1-u_1\partial^{-1}(q+\omega_1)\nonumber\\
M_{41}&=&(r+\omega_2+u_2)\partial^{-1}u_2+u_1\partial^{-1}(r+\omega_2)\nonumber\\
M_{42}&=&-(r+\omega_2+u_2)\partial^{-1}u_1-u_2\partial^{-1}(q+\omega_1)\nonumber\\
M_{43}&=&(r+\omega_2+u_2)\partial^{-1}(r+\omega_2+u_2)\nonumber\\
M_{44}&=&\frac{1}{2}\partial-(r+u_2+\omega_2)\partial^{-1}(q+\omega_1+u_1)\nonumber\\
M_{47}&=&(r+\omega_2+u_2)\partial^{-1}u_2+u_2\partial^{-1}(r+\omega_2)\nonumber\\
M_{48}&=&-(r+\omega_2+u_2)\partial^{-1}u_1-u_2\partial^{-1}(q+\omega_1)\nonumber\\
M_{51}&=&(q-\omega_1+v_1)\partial^{-1}v_2+v_1\partial^{-1}(r+\omega_2)\nonumber\\
M_{52}&=&-(q-\omega_1+v_1)\partial^{-1}v_1-v_1\partial^{-1}(q-\omega_1)\nonumber\\
M_{55}&=&-\frac{1}{2}\partial+(q+v_1-\omega_1)\partial^{-1}(r-\omega_2+v_2)\nonumber\\
M_{56}&=&-(q-\omega_1+v_1)\partial^{-1}(q-\omega_1+v_1)\nonumber\\
M_{57}&=&v_1\partial^{-1}(\omega_2-r)-(q+v_1-\omega_1)\partial^{-1}v_2\nonumber\\
M_{58}&=&-v_1\partial^{-1}(\omega_1-q)+(q+v_1-\omega_1)\partial^{-1}v_1\nonumber\\
M_{61}&=&v_2\partial^{-1}(\omega_2-r)+(r+v_2-\omega_2)\partial^{-1}v_2\nonumber\\
M_{62}&=&v_2\partial^{-1}(\omega_1-q)-(r+v_2-\omega_2)\partial^{-1}v_1\nonumber\\
M_{65}&=&(r-\omega_2+v_2)\partial^{-1}(r-\omega_2+v_2)\nonumber\\
M_{66}&=&\frac{1}{2}\partial-(r+v_2-\omega_2)\partial^{-1}(q-\omega_1+v_1)\nonumber\\
M_{67}&=&-v_2\partial^{-1}(\omega_2-r)-(r+v_2-\omega_2)\partial^{-1}v_2\nonumber\\
M_{68}&=&v_2\partial^{-1}(\omega_1-q)-(r+v_2-\omega_2)\partial^{-1}v_1\nonumber\\
M_{71}&=&-q\partial^{-1}\omega_2-\omega_1\partial^{-1}r\nonumber\\
M_{72}&=&q\partial^{-1}\omega_1+\omega_1\partial^{-1}q\nonumber\\
M_{78}&=&q\partial^{-1}q+\omega_1\partial^{-1}\omega_1\nonumber\\
M_{81}&=&r\partial^{-1}\omega_2+\omega_2\partial^{-1}r\nonumber\\
M_{82}&=&-r\partial^{-1}\omega_1-\omega_2\partial^{-1}q\nonumber\\
M_{87}&=&r\partial^{-1}r+\omega_2\partial^{-1}\omega_2\nonumber\\
M_{88}&=&\frac{1}{2}\partial-r\partial^{-1}q-\omega_2\partial^{-1}\omega_1\nonumber\\
\end{eqnarray}
and $M_{ij}=0$ for $(i=1,2;j=3,4,5,6)$, $(i=3,4,j=5,6)$ and  $(i=7,8;j=3,4,5,6)$
Corresponding coupled equations can also be obtained as before.

\section{References:}

\par[1] B. Fuchssteiner-Prog. Theor. Phys. \textbf{65}(1981)861 and Prog. Theor. Phys. \textbf{70}(1983)1508.\\

\par [2]B. Fuchssteiner and A. S. Fokas-Physica D  \textbf{4}(1981)47.\\

\par [3]Lie Algebraic Methods In Integrable System-A. RoyChowdhury(Chapman and Hall, CRC Prass)\\

\par [4] Y. Zhang-Phys. Lett. A.\textbf{317}(2003)280.\\

\par [5] Bacicai and Gu Zhuqan-J. Phys. A \textbf{24}(1991)963.\\

\par [6]Zhang and H. Tau-Chaos, Solitons, Fractals \textbf{39}(2009)1109.\\

\par [7] F. B. Wahlquist and H. D. Estabrook-J. Math. Phys.\textbf{17}(1976)1293. \\

\par [8]A. RoyChowdhury and S. Ahamed-Phys. Rev. D\textbf{10}(1985)2780.\\

\par [9]Yu Fa Jun-Chin. Phys B  \textbf{21}(2012)(010201).\\

\par[10] Binlu Feng and Yu Feng Zhang-Advances in Math. Phys.\textbf{2015B}(2015)154915.\\

\par[11] Solitons- Springer-Verlag, Berlin edited by Dodd and R. K. Bullough.\\

\par[12] Li Li and Fajun Yu-Int. J. Nonlinear Sci. Num. Simul\textbf{14}(2013)513.\\

\par[13] B. Fenf-J. Phys. A. Theor. Math.\textbf{45}(2012)085202.\\

\par[10] G. Tu-J. Math. Phys. \textbf{30}(1989)330.\\
\end{document}